\documentclass[proceedings]{JHEP3}

\PrHEP{PrHEP hep2001}                   
\conference{International Europhysics Conference on HEP}                

\usepackage{epsfig}                   

\title{New Results on ATLAS Pixel Opto-Link}

\author{\speaker{K.K. Gan}\\
        Department of Physics,
        The Ohio State University,
        Columbus, OH 43210,
        U.S.A.\\
        E-mail: \email{gan@mps.ohio-state.edu}}

\abstract
{We present new results on the optical link for the pixel detector of the ATLAS experiment.
An optical package of novel design has been developed for the opto-link.
The design is based on a simple connector-type concept and is made of
radiation-hard material.
The receiver (DORIC) and transmitter (VDC) chips have been designed.
The prototype results using the 0.8 and 0.25 $\mu$m technologies are presented.}

\begin{document}

The ATLAS pixel detector~\cite{pixel} consists of three barrel layers and
three forward and backward disks.
The detector covers the pseudo-rapidity region $| \eta | < 2.5$ and
provides at least three space point measurements.
The low voltage differential signal (LVDS) from the pixel detector is converted
by the VCSEL Driver Chip (VDC) into a single-ended signal appropriate to drive
a Vertical Cavity Surface Emitting Laser (VCSEL).
The optical signal is transmitted to the Readout Device (ROD) via a fibre.
For the innermost barrel layer, the signal is transmitted using two fibres.
The 40-MHz beam crossing clock from the ROD, bi-phase encoded with the command
signal to control the pixel detector, is transmitted to a PIN diode via a fibre.
The PIN signal is decoded using a Digital Opto-Receiver Integrated Circuit (DORIC).
The VSCEL and PIN diodes couple to the fibres inside optical packages.
In this paper, we describe the performance of the opto-pack developed by The
Ohio State University (OSU) group and status of the VDC and DORIC prototypes.

\section{Results on Opto-Packs}

The main technical challenge in the fabrication of the opto-pack is
the alignment tolerance of the VCSEL with respect to the fibre:

\hspace{0.1in} $\bullet$ 50 $\mu$m in $z$ (along the fibre)

\hspace{0.1in} $\bullet$ 25 $\mu$m in $r$ (transverse to the fibre)

\noindent The requirements can be satisfied either by passive or
active alignment.
For the former, parts must be fabricated or placed with high precision
($\le 10~\mu$m) so that the overall precision is still within the tolerance.
The alignment tolerance of the PIN to the fiber is much looser.
There were two designs for the opto-pack, OSU and Academia Sinica (Taiwan),
using passive and active alignment, respectively.

In the Taiwan design, the fiber is cleaved at 45$\rm ^o$ to act
as a mirror.
Light traveling down the fiber is reflected onto the PIN
placed directly below the cleaved surface.
Conversely, light emitted by a VCSEL is reflected off the
cleaved surface into the fiber.
Each fiber is actively aligned and glued permanently to the package.
The package is of low cost but the permanent attachment of the
fibers to the package presents a technical challenge to the
assembly of the pixel detector.

\EPSFIGURE{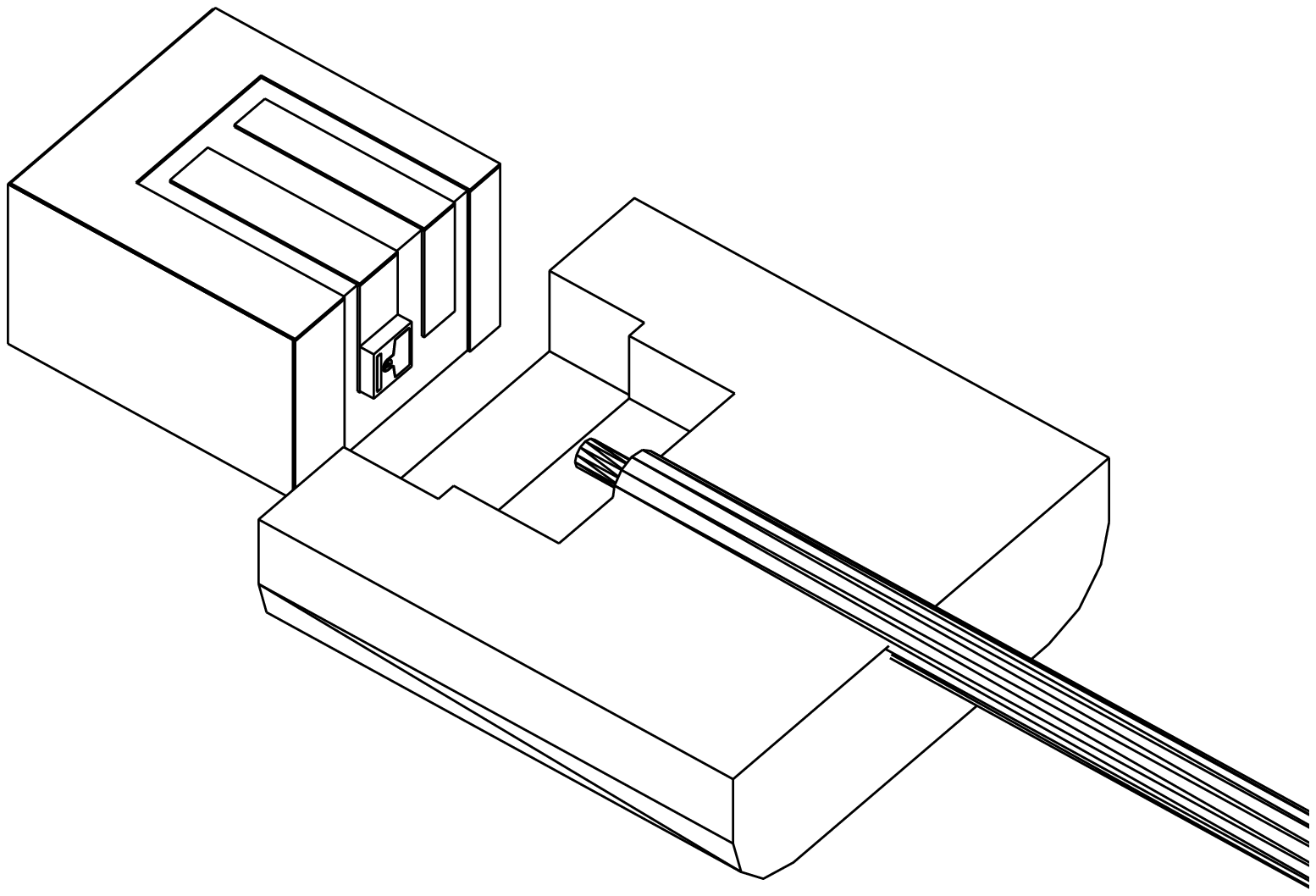,width=4.0cm}                       
{A cut out view of the opto-pack showing a cap plus a base with a VCSEL mounted.
\label{cap_base_fibre}}

The OSU design uses a connector concept: a cap with a hole for the fibre and
a base with deposited wire bonding trace and pad for PIN or VCSEL placement.
The design is shown in Fig.~\ref{cap_base_fibre}.
The package is of low cost and the cap with the fiber attached can be
mounted on the base near the end of the pixel detector assembly.
Precise alignment of the VCSEL to the fiber is achieved by fabricating
the bases and caps with high precision and by the accurate placement
of the VCSEL and fibre relative to the base and cap, respectively.

The base is made of aluminia.
To fabricate the base, aluminia sheet is ground to the precise
thickness of the base and then cut into strips for deposition
of three-dimensional traces~\cite{HT}.
Most of the deposited traces have good connectivity across
the corner of the base.
Strips with a large number of traces of good connectivity
were then precisely diced into individual bases~\cite{AD}.

The cap is made of Ultem (polyetherimide), a mold-injectable plastic
with a radiation tolerance~\cite{rad} of 10 GRad.
Due to the high cost and long lead time in developing the 
injection molding technique for fabricating small parts with high precision,
we decided to use ``manual mold injection''.
Here a small spring-loaded mold of 5~cm $\times$ 5~cm $\times$ 10~cm
is used in a small oven as opposed to a standard mold of
30~cm $\times$ 30~cm $\times$ 30~cm placed inside an automatic mold
injection machine of 1~m $\times$ 1~m $\times$ 2~m.
The critical part in the mold fabrication is the precisely machined mold
for the interior of the cap that fits the base and the precisely located
hole for the placement of a pin which produces the hole for the fiber.
We have proven the principle of this precise micro-mold injection
technique and can fabricate several quality caps per hour.

The VCSEL is precisely placed on the base under an optical comparator.
To further improve the alignment between the fiber and the VCSEL,
the location of the pin on the mold is also measured so that the VCSEL
is placed on the base at the location as expected from the fabricated cap.
In addition, the dimension of each base is measured individually so that
the location of the VCSEL is adjusted slightly to account for the small
variation in size.

We have fabricated 31 VCSEL opto-packs.
Figure~\ref{power_base} shows the measured coupled power.
The coupled power of all packages are above the specification of
300~$\mu$W minimum power for an average current of 10~mA in the VCSEL.
The average coupled power is 760~$\mu$W.
The caps can also be used multiple times.
Figure~\ref{power_trial} shows the coupled power of two typical caps
for each of the 10 mountings on a base.
Each mounting gives about the same coupled power.
The caps are also interchangeable.
Figure~\ref{power_cap} shows the coupled power of a cap on the 31 bases.
The cap gives about the same coupled power on all bases.
This proves the feasibility of the connector-type design.
The waveforms of the mounted VCSELs also have fast rise time, $<$ 1ns.
We have fabricated two PIN opto-packs with three PINs in each package.
The PIN diodes have good responsivity, $>$ 0.5~A/W.
The opto-packs therefore meets the pixel detector specification.

\EPSFIGURE{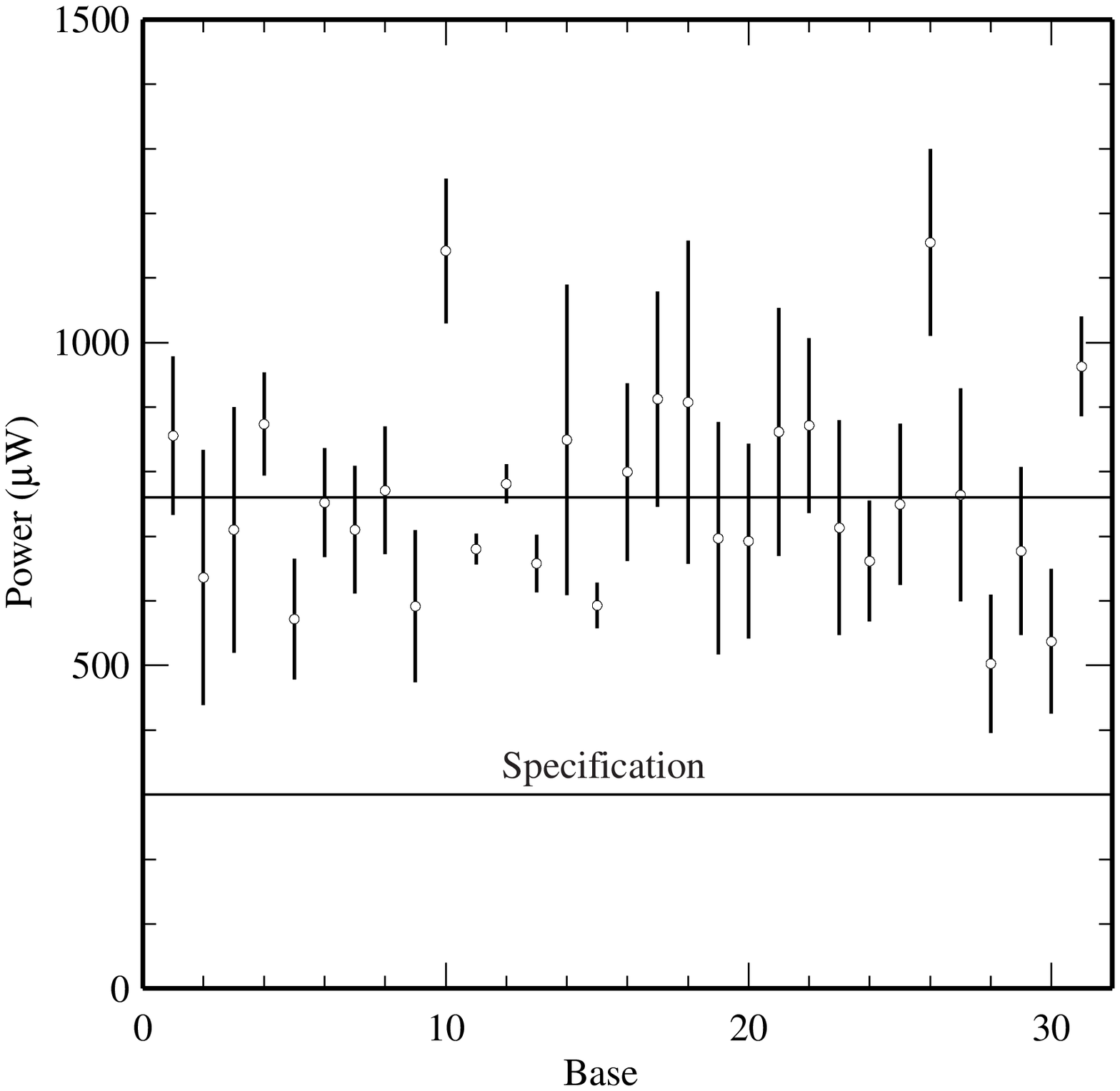,width=5.0cm}                       
{Measured coupled power of 31 VCSEL bases.
The error bars correspond to the standard deviations of measurements from
about five caps on each base.
\label{power_base}}

There was a review of the OSU and Taiwan opto-packs in June.
The Taiwan opto-pack potentially has 10-15\% higher coupled
power and has been chosen as the baseline in order to take
advantage of the additional resources from Taiwan.

\section {Results on Optical Chips}

The pixel detector design of the VDC and DORIC takes advantage of the development
work for similar chips used by  the outer detector, the SemiConductor Tracker (SCT).
Both SCT chips attain radiation-tolerance by using bipolar integrated circuits
(AMS 0.8~$\mu$m BICMOS) and running with high currents in the transistors.
The chip is therefore not applicable for the higher radiation dosage
and lower power budget requirements of the pixel detector.
We design the radiation-hard CMOS version of the circuits
in collaboration~\cite{team} with University of Siegen, Germany.

\EPSFIGURE{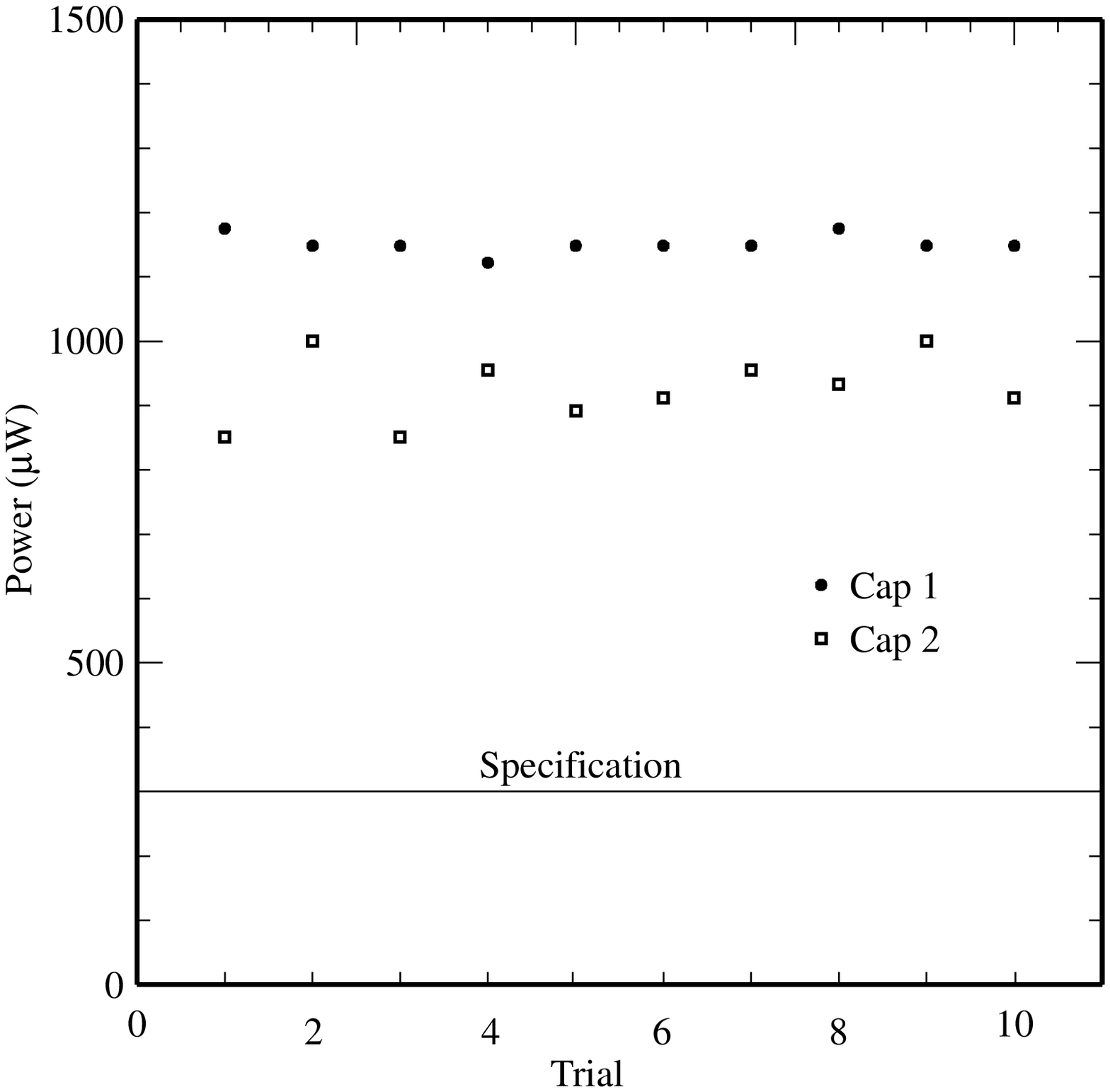,width=5.0cm}                       
{Measured coupled power with two different caps on a VCSEL base.
\label{power_trial}}

The first pixel version of the chips with 0.8~$\mu$m feature size
was submitted to DMILL in July 1999.
VDC functioned well but DORIC could not decode the clock properly.
The problem was due to underestimate of the parasitic capacitance.

The second version of the chips was submitted to DMILL in July 2000.
Both chips worked but for some of the DORICs the pre-amp output was observed
to be saturated with no input signal, indicating that there was an offset
between the differential pre-amp inputs due to process variations.
We submitted to DMILL in May 2001 a new design with a DC feedback
circuit between the output and input of the pre-amp to cancel the offset.
We expect the delivery of the chips in November 2001.

In April 2001, we irradiated 14 chips from the second DMILL run with
24~GeV protons at CERN.
Some of the chips can operate up to $\sim$~50~Mrad by increasing
the supply voltage from 3.2 to 5.0~V but others died within a few Mrad.
Annealing at 100$\rm ^o$C for a week failed to improve or revive the chips.
This indicates that the chips may not be adequate for the pixel detector
as we expect the optical link to receive a dosage of 20~Mrad.

\EPSFIGURE{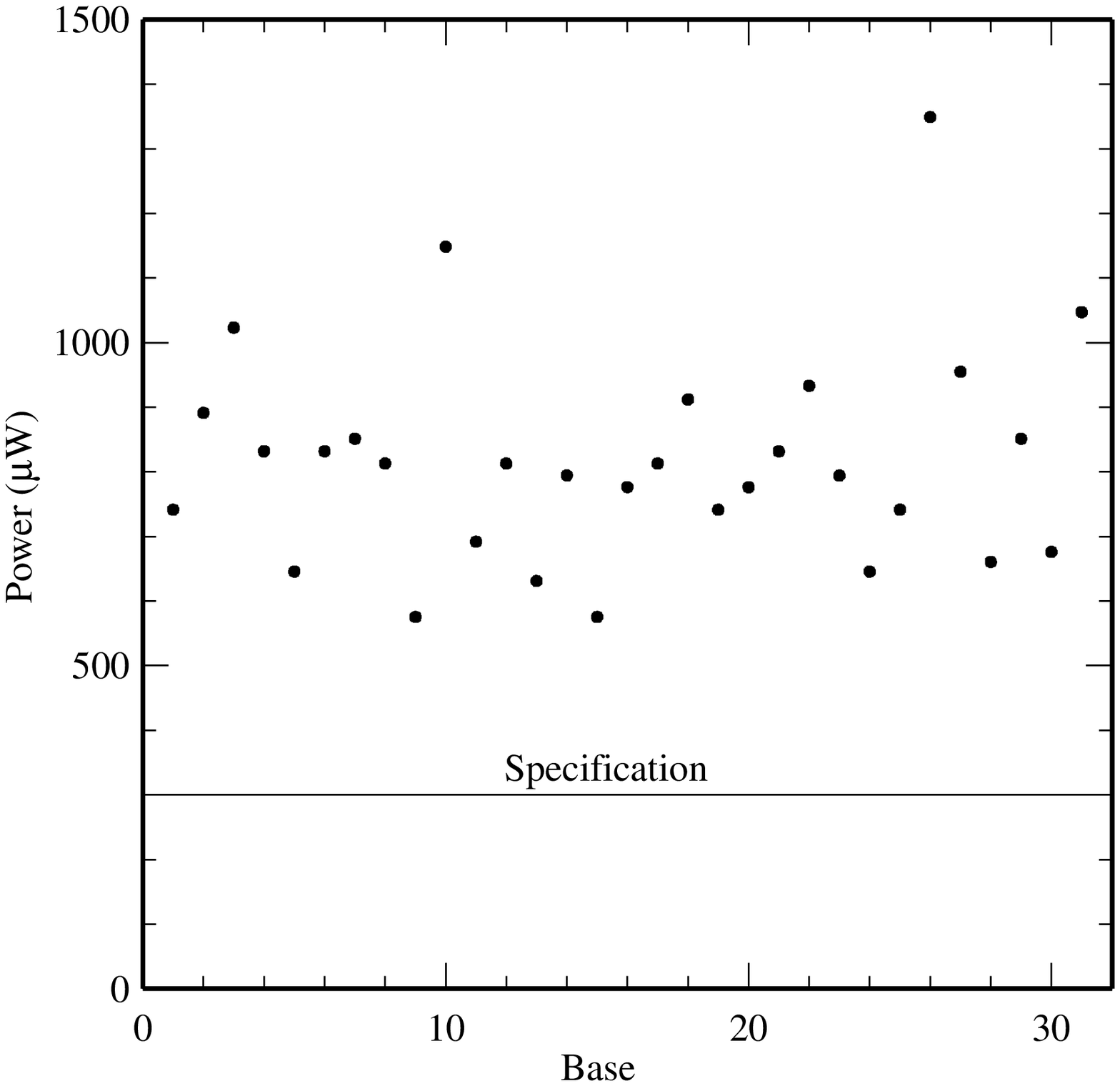,width=5.0cm}                       
{Measured coupled power of 31 VCSEL bases with the same cap.
\label{power_cap}}

We have converted the DMILL design into
a design with smaller feature size, 0.25~$\mu$m, using a different vendor.
This process is expected to be more radiation hard than the DMILL process.
The new designs were submitted in February 2001.
The VDC works well with fast rise and fall time ($<$ 1 ns) but consumes more
current than we would like.
We plan to reduce the current consumption by several mA in the next submission
expected in the Fall 2001.
For the DORIC, the delay control circuit oscillates.
We are not able to reproduce the oscillation in the simulation.
The oscillation can be cured with an external bypass capacitor and
then the chips decode the bi-phase marked clock properly.
Some chips require a large input PIN signal due to the large pre-amp offset from process variations.
The DC feedback circuit can cancel part of the offset; this has been verified
by adjusting the amount of DC feedback.
We will increase the amount of feedback in the next submission.
We plan to irradiate the chips with 24~GeV protons at CERN in September 2001.

\section{Summary}
We have developed an opto-pack based on the novel connector-type concept for
the pixel detector of the ATLAS experiment.
The opto-pack meets the specification but the Taiwan design was chosen as the
baseline to take advantage of their resources.
We have designed  versions of the VDC and DORIC in the 0.8 and 0.25~$\mu$m technology.
The chips are functional but further improvements are planned.
The design in the 0.8~$\mu$m technology appears inadequate in radiation
hardness for the ATLAS pixel detector.
\acknowledgments
This work was supported in part by the U.S.~Department of Energy.
The author wishes to thank K.E. Arms, K. Arndt, J. Burns, H.P. Kagan, R.D.~Kass, S. Smith,
T. Weidberg, and R. Wells for their contributions to the development of the opto-packs.



\begin{thebibliography}{99}
\bibitem{pixel} ATLAS Pixel Detector Technical Design Report,
         CERN/LHCC/98-13.

\bibitem{HT} Hybrid-Tek Inc., 1 Hytek Corporate Ctr, Rte. 526, Clarksburg, NJ 08510, USA.

\bibitem{AD} American Dicing Inc., 344 East Brighton Ave., Syracuse, NY 13210, USA.

\bibitem{rad} M.~Tavlet, A.~Fontaine, and H.~Schonbacher, Compilation of Radiation
        Damage Test Data, CERN Report No. CERN 98-01, 1998.

\bibitem{team} The optical electronics team members from The Ohio State University
        are	K.E. Arms, K.K. Gan, M.O. Johnson, H.P. Kagan,
		      R.D. Kass, C. Rush, and M. Zoeller.   The team members from
        University of Siegen are	M. Kraemer, J. Hausmann, M. Holder,
		      and M. Ziolkowski.
\end{thebibliography}
\end{document}